\documentclass[twocolumn]{autart}
\pdfoutput=1
\usepackage[utf8]{inputenc}
\usepackage{hyperref}
\usepackage{amsmath,amssymb}
\usepackage[english]{babel}
\usepackage[table]{xcolor}
\usepackage{tikz,pgfplots}
\usetikzlibrary{intersections, backgrounds, positioning}

\makeatletter
\def\@biblabel#1{}
\makeatother

\usepackage[round]{natbib}

\makeatletter
\newcommand{\pushright}[1]{\ifmeasuring@#1\else\omit\hfill$\displaystyle#1$\fi\ignorespaces}
\newcommand{\pushleft}[1]{\ifmeasuring@#1\else\omit$\displaystyle#1$\hfill\fi\ignorespaces}
\makeatother

\newcommand{\T}{\mathsf{T}}
\renewcommand{\H}{\mathsf{H}}
\newcommand{\R}{\mathbb{R}}
\newcommand{\C}{\mathbb{C}}

\newcommand{\ri}{\mathsf{i}}
\newcommand{\dd}{\mathsf{d}}
\DeclareMathOperator{\errmeasure}{\mathcal{E}}

\newcommand{\dimx}{n}
\newcommand{\dimu}{m}
\newcommand{\dimy}{p}
\newcommand{\pHsys}{\Sigma_{\mathsf{pH}}}
\newcommand{\nSamples}{n_s}

\newcommand{\tfPHFit}{H_{\mathsf{pH}}}
\newcommand{\Hstorage}{\mathcal{H}_{\mathsf{st}}}

\newcommand{\fullE}{E_{\mathsf{g}}}
\newcommand{\fullA}{A_{\mathsf{g}}}
\newcommand{\fullB}{B_{\mathsf{g}}}
\newcommand{\fullC}{C_{\mathsf{g}}}
\newcommand{\fullD}{D_{\mathsf{g}}}

\newcommand{\vfFun}{H_\mathsf{vf}}
\newcommand{\vfC}{C_{\mathsf{vf}}}
\newcommand{\vfB}{B_{\mathsf{vf}}}
\newcommand{\vfA}{A_{\mathsf{vf}}}
\newcommand{\vfE}{F_{\mathsf{vf}}}
\newcommand{\vfD}{D_{\mathsf{vf}}}

\DeclareMathOperator{\vtf}{vtf}

\DeclareMathOperator{\vtu}{vtu}
\DeclareMathOperator{\utv}{utv}
\DeclareMathOperator{\vectorize}{vec}
\DeclareMathOperator{\vtsu}{vtsu}
\DeclareMathOperator{\sutv}{sutv}
\DeclareMathOperator{\Real}{Re}
\renewcommand{\hat}{\widehat}

\newcommand{\uu}{\widehat u}
\newcommand{\vv}{\widehat v}
\newcommand{\dyn}{\mathcal{D}_0}
\newcommand{\dyninvinp}{a}
\newcommand{\outinvdyn}{b}

\newcommand{\dE}{d\theta_E}
\newcommand{\dJ}{d\theta_J}
\newcommand{\dW}{d\theta_W}
\newcommand{\dB}{d\theta_B}
\newcommand{\dN}{d\theta_N}

\DeclareMathOperator{\trace}{tr}

\newcommand{\loss}{\mathsf{L}}

\begin{document}

\begin{frontmatter}
  \title{Port-Hamiltonian System Identification from Noisy Frequency Response Data}
  \thanks[footnoteinfo]{This work is supported by the German Research Foundation (DFG) within the project VO2243/2-1: ``Interpolationsbasierte numerische Algorithmen in der robusten Regelung"}
  \author[TUB]{Paul Schwerdtner}
  \address[TUB]{Technische Universit\"at Berlin, Institut f\"ur Mathematik, Stra{\ss}e des 17. Juni 136, 10623 Berlin, Germany. \texttt{schwerdt@math.tu-berlin.de}}
  \begin{abstract}
    We present a new method for the identification of linear time-invariant passive systems from noisy frequency response data. In particular, we propose to fit a parametrized port-Hamiltonian (pH) system, which is automatically passive, to supplied data with respect to a least-squares objective function.
    In a numerical study, we assess the accuracy of the resulting identified models by comparing our method to two other frequency domain system identification methods. One of the methods being compared is a recently published identification procedure that also computes pH systems and the other one is the well-known vector-fitting algorithm, which provides unstructured models. The numerical evaluation demonstrates a substantial increase in accuracy of our method compared to the other pH identification procedure and a slightly improved accuracy compared to vector-fitting. This underlines the suitability of our method for the estimation of passive or pH systems -- in particular from noisy frequency response data.
  \end{abstract}
\end{frontmatter}

\section{Introduction}
We study the identification of \emph{passive} linear time-invariant (LTI) dynamical systems from noisy frequency response data (FRD).
Consider an LTI system in the generalized state-space form given by
\begin{align}
  \Sigma:\,
  \begin{cases}
    \fullE \dot x(t) = \fullA x(t)+\fullB u(t), \\
    \phantom{\fullE} y(t) = \fullC x(t) +\fullD u(t).
  \end{cases}
  \label{eq:std_sys}
\end{align}
Here, $\fullE,\fullA \in \R^{\dimx \times  \dimx}$, $\fullB \in \R^{\dimx \times  \dimu}$, $\fullC \in \R^{\dimy \times \dimx}$, and $\fullD \in \R^{\dimy \times  \dimu}$ are the system matrices defining $\Sigma$. We call $x: \R  \rightarrow \R^{\dimx}$, $u : \R  \rightarrow \R^{\dimu}$, and $y: \R  \rightarrow \R^{\dimy}$ the \emph{state}, \emph{input}, and \emph{output} of $\Sigma$, respectively, and assume that $x(0) = 0$.
Applying the \emph{Laplace} transform to both equations in \eqref{eq:std_sys} and eliminating the state yields the \emph{transfer function} of $\Sigma$ given by
\begin{align}
  \label{eq:StdTF}
   H(s) := \fullC (s\fullE-\fullA)^{-1}\fullB+\fullD.
\end{align}
The \emph{frequency response} of $\Sigma$ at a given frequency $\omega \in \R$ is then given by $H(\ri \omega)$, where $\ri$ denotes the imaginary unit. Conversely, $\Sigma$ is also called a \emph{realization} of $H$.

The frequency domain system identification task is to estimate a system $\Sigma_{\mathsf{id}}$ based on $n_s \in \mathbb{N}$ samples of FRD given as $(\hat H_i, \ri \omega_i)_{i=1}^{\nSamples}$, with $\hat H_i \in \C^{\dimy \times \dimu}$ and $\omega_i \in \R$. 
In this way, system identification is related to data-driven model order reduction, in which a realization with a small state dimension is constructed based on evaluations of the transfer function of a given model with a large state dimension. In model order reduction, the FRD can thus simply be obtained by evaluating the transfer function of a known LTI system. However, in this work, we particularly address the case when the transfer function \emph{cannot} be directly accessed and instead the FRD must be obtained indirectly from either running simulations of a complex dynamical model or conducting experiments on an existing plant by exploiting the well-known relationship between time-domain data and FRD (\cite{Ljung1985} and \cite{Peh2017}). Thus, the identification method must be able to handle FRD that is corrupted by numerical or measurement noise.

Furthermore, we construct an identification method which generates a passive system. Passivity is a useful tool in the domain of networked systems, as the power-conserving interconnection of passive systems is again passive and the passivity of a system implies its stability. With increasing deployment of passive systems obtained from first principle modeling enabled by recent developments in \cite{Egger2019} and \cite{Serhani2019}, also the identification of passive systems from data becomes a pressing issue.
This is because only such identification methods enable the construction of hybrid networked systems consisting of passive sub-systems obtained via both modeling and identification. We refer to \cite{Wil1972} for a detailed survey on passive systems and their properties and only briefly summarize the prerequisites necessary to describe our method.

A system $\Sigma$ is passive if there exists a state-dependent energy storage function $\Hstorage(x(t)) \ge 0$, such that the \emph{dissipation inequality}
\begin{align*}
  \Hstorage(x(t_1)) - \Hstorage(x(t_0)) \le \int\limits_{t_0}^{t_1}\Real\left(y(t)^\H u(t)\right) \dd t,
\end{align*}
holds for all $t_1, t_0 \in \R$ with $t_1>t_0$. A passive system, that is also minimal, has a positive real transfer function \cite[Theorem~1]{Wil1972}, as defined in the following.
\begin{defn} (Positive real transfer functions) \\
  A proper rational transfer function $H$ is called positive real if
  \begin{itemize}
    \item all poles of H have non-positive real part, 
    \item the matrix valued function
      \begin{align}
        \label{eq:spectral}
        \Phi(s) := H(-s)^{\T} + H(s)
      \end{align}
      attains positive semi-definite values for all $s \in \ri \R$, which are not poles of $H$,
    \item for any purely imaginary pole $\ri \omega$ of $H$ we have that the residue matrix $\lim \limits_{s  \rightarrow \ri \omega}(s-\ri \omega)H(s)$ is positive semi-definite.
  \end{itemize}
  The zeros of $\Phi$, i.e.\ all points $s \in \C$, where $\det(\Phi(s)) = 0$, are called \emph{spectral zeros} of $H$.
\end{defn}

Hence, a passivity enforcing frequency domain identification method must perform the following task. Given potentially noisy FRD, a positive real transfer function $H_{\mathsf{id}}$ and its realization must be \emph{estimated}. Existing methods either provide positive real interpolants of given FRD, which does not perform well for noisy data, or fail to generically identify a passive system and instead only guarantee that the identified system is stable. On the other hand, we propose a method that handles noise well and guarantees the passivity of identified systems. The method is based on directly optimizing the parameters of a passive parametric system to obtain a least-squares fit to the given FRD.

\subsection{Port-Hamiltonian realizations of passive systems}

We make use of port-Hamiltonian (pH) realizations, which are a special class of realizations of passive systems. In fact, it has been shown in \cite{Beattie2016} that passive systems have a pH realization. In \cite{Beattie2019} and \cite{Meh2020OptRobustness} the robustness of pH realizations is investigated and it is found that a certain pH realization of a passive system maximizes the passivity radius. Throughout this work, we employ generalized pH systems based on \cite{Beattie2018}.

\begin{defn} (Generalized Port-Hamiltonian systems) \\
  \label{def:pHDesSys}
  An LTI system of the form
  \begin{align}
    \pHsys: \begin{cases}
      E\dot x(t)= (J-R)x(t) + (B-P) u(t), \\
      \phantom{E}y(t)= (B+P)^\T x(t) + (S+N) u(t),
    \end{cases}
  \end{align}
  with $E, J, R \in \R^{\dimx \times \dimx}$, $B,P \in \R^{\dimx \times \dimu}$, and $S,N \in \R^{\dimu \times \dimu}$ is called a port-Hamiltonian system, if $E \ge 0$, $J=-J^\T$, $N=-N^\T$, and
  \begin{align}
    W:=
    \begin{bmatrix}
      R & P \\ P^\T & S
    \end{bmatrix}
    \ge 0.
  \end{align}
  Its \emph{Hamiltonian} (energy storage) function is given by
  \begin{align*}
    \mathcal{H}(x(t)) = \frac{1}{2}x(t)^{\T}E x(t).
  \end{align*}
\end{defn}

In particular, we propose to fit a model in generalized pH system format to given FRD to address the frequency domain identification task.

\subsection{Related Work}

In \cite{Ben2020} an approach for determining generalized pH systems based on FRD is proposed, in which the authors use the Loewner matrix framework proposed in \cite{AntM07} and further explained in \cite{AntLI17} to construct a passive realization, which interpolates given FRD. This extends the work of \cite{Ionutiu2008} to the domain of system identification. The procedure proposed in \cite{Ben2020} consists of two steps. In a first step, a conventional rational interpolant $Z$ (with the same structure as in \eqref{eq:StdTF}) is constructed for given FRD using the Loewner framework. Then the spectral zeros of~$Z$ are computed and used as interpolation data for yet another Loewner based realization. The construction of that second realization as defined in \cite[Alg.~1]{Ben2020} guarantees passivity. Furthermore, this second realization is in pH form. Throughout this work we refer to this method as pH-Loewner.

The state dimension of the Loewner interpolants is $\nSamples \cdot m$, which is too large for most applications, since usually a high number of frequency samples is available. However, for noise free FRD the state dimension can often be reduced to a moderate size while preserving the pH structure exploiting the typically rapid decay of the singular values of the constructed Loewner matrix pencil. Problems occur when the FRD is corrupted by noise. In this case, interpolation of the FRD samples without any regularization naturally does not lead to a good overall fit, which we also observe in our numerical experiments (consider e.g.\ Figure~\ref{fig:mean_errors}). Moreover, the constructed surrogate models cannot be reduced anymore, since for noisy interpolation data, the singular values of the Loewner matrix pencils do not decay as rapidly as in the noise free case.

On the other hand, in the well-established vector-fitting-framework for rational realizations initially proposed in \cite{Gus99} and refined in \cite{Gus06} and \cite{Des08} noisy FRD is handled well. Vector-fitting is based on optimizing poles and residues of a rational transfer function in the format
\begin{align}
  \vfFun(s) := \vfC(sI_{\dimx}-\vfA)^{-1}\vfB+\vfD+s\vfE,
  \label{eq:vfFunction}
\end{align}
to fit $\vfFun$ to given FRD. Here $I_{\dimx}$ denotes the identity matrix in $\C^{\dimx \times \dimx}$, $\vfC^\T, \vfB \in \C^{\dimx \times \dimu}$, $\vfA\in \C^{\dimx \times \dimx}$, and $\vfD, \vfE \in \C^{\dimu \times \dimu}$. The system order $\dimx$ can be chosen beforehand and independently of $\nSamples$. In vector-fitting, the minimized objective function is of a least-squares type, which leads to the suitability of this method to noisy FRD. Note that both $\vfE$ and $\vfD$ can be constrained to zero in case $\vfFun$ is required to be proper or strictly proper, respectively. However, even if the FRD is generated by a passive system, the surrogate model returned by the vector-fitting algorithm is not guaranteed to be passive and thus cannot be formulated as a pH model in general.

\begin{rem}
  The transfer function \emph{$\vfFun$} can also be realized as generalized state-space model, such that it also has a representation as in \eqref{eq:StdTF}. However, we introduce \emph{$\vfFun$} as in \eqref{eq:vfFunction} since this is the format used in the vector-fitting algorithm.
\end{rem}

We also mention the AAA algorithm (\cite{Nak2018}), which allows to approximate rational functions using a combination of greedy interpolation and least-squares fitting, and \cite{Cherif2019}, in which several \emph{indirect} methods for the construction of pH systems from time-domain data are presented. However, neither the AAA algorithm nor the indirect methods for the construction of pH systems can directly be utilized for pH system identification from noisy data. Therefore, we only consider pH-Loewner and vector-fitting in a comparison with our proposed method to evaluate the accuracy. For an overview of previous general purpose frequency domain system identification methods we refer to \cite{Pin1994}.

\subsection{Outline}

In what follows, we extend recent work (\cite{SchV20}), in which a passivity preserving model order reduction method for strictly proper passive systems is proposed. This method is data driven in the sense that it solely uses evaluations of the transfer function of a given model to compute a low order (and also passive) surrogate model and no system matrices of the given model are directly accessed. Instead, the method is based on fully parametrized pH systems. The parametrization is optimized in the course of the model order reduction procedure such that its transfer function approximates the transfer function of the given model in the $\mathcal{H}_\infty$ norm.

In the next section, we explain how this model order reduction method can be adapted for the system identification task. For that we generalize the parametrization from \cite{SchV20} for systems as defined in Def.~\ref{def:pHDesSys} and provide details for solving the parameter optimization problem. We investigate the accuracy of the identified systems and the influence of noise on the identification results in Section~\ref{sec:numerical_results}.

\section{Port-Hamiltonian System Parametrizations}

Following the ideas of \cite{SchV20}, we design the parametrization of $\pHsys$ such that for all parameter vectors $\theta \in R^{n_\theta}$, the structural constraints imposed in Def.~\ref{def:pHDesSys} are automatically satisfied. This permits the usage of unconstrained optimization solvers during identification. We rely on the following functions to construct the system matrices.

The family of functions
\begin{align*}
  \vtf_m: \C^{n \cdot m} \rightarrow \C^{n\times m}, \;  v \mapsto
  \begin{bmatrix}
    v_1 & v_{n+1} & \dots & v_{m(n-1)+1}\\
    v_2 & v_{n+2} & \dots & v_{m(n-1)+2}\\
    \vdots & \vdots &  & \vdots \\
    v_n & v_{2n} & \dots & v_{nm}
  \end{bmatrix}
\end{align*}
reshapes a vector into an accordingly sized matrix with $m$ columns.
Its inverse is the standard vectorization operator denoted by $\vectorize$.
The function 
\begin{align*}
  \vtu : \C^{n(n+1)/2}  \rightarrow \C^{n\times n}, \; v \mapsto
  \begin{bmatrix}
    v_1 & v_2 & \dots & v_n \\
    0 & v_{n+1}& \dots & v_{2n-1} \\
    0 & 0 & \ddots & \vdots \\
    0 & 0 & 0 & v_{n(n+1)/2} \\
  \end{bmatrix}
\end{align*}
maps a vector of length $n(n+1)/2$ to an $n\times n$ upper triangular 
matrix, while the function $\utv : \C^{n\times n}  \rightarrow \C^{n(n+1)/2}$
maps the upper triangular part of a given matrix row-wise to a vector.
The function
\begin{align*}
  \vtsu : \C^{n(n-1)/2}  \rightarrow \C^{n\times n}, \; v \mapsto
  \begin{bmatrix}
    0 & v_1 & v_2 & \dots  & v_{n-1} \\
    0 & 0   & v_n & \dots  & v_{2n-2}  \\
    0 & 0   & 0   & \ddots & \vdots \\
    0 & 0   & 0   & 0      & v_{n(n-1)/2} \\
    0 & 0   & 0   & 0      & 0\\
  \end{bmatrix}
\end{align*}
maps a vector of length $n(n-1)/2$ to an $n\times n$ strictly upper 
triangular matrix, while the function $ \sutv : \C^{n \times n}  \rightarrow \C^{n(n-1)/2} $
maps the strictly upper triangular part of a given matrix row-wise to a 
vector.

With these operations, we can define a parametrization of generalized pH systems as follows.

\begin{lem}
  \label{lem:PHParam}
  Let $\theta \in \R^{n_\theta}$ be a parameter vector partitioned as $\theta = \begin{bmatrix}
    \theta_E^\T, & \theta_J^\T, & \theta_W^\T, & \theta_B^\T, & \theta_N^\T
  \end{bmatrix}^\T$, with $\theta_E \in \R^{\dimx(\dimx+1)/2}$, $\theta_J \in \R^{\dimx(\dimx-1)/2}$, $\theta_W \in \R^{(\dimx+\dimu)(\dimx+\dimu+1)/2}$, $\theta_B \in \R^{\dimx \cdot \dimu}$, and $\theta_N \in \R^{\dimu(\dimu-1)/2}$.
  Furthermore, define the matrix valued functions
  \begin{subequations}
    \begin{align}
      E(\theta) &:= \vtu(\theta_E)^\T \vtu(\theta_E),\\
      J(\theta) &:= \vtsu(\theta_J)^\T - \vtsu(\theta_J),\\
      W(\theta) &:= \vtu(\theta_W)^\T \vtu(\theta_W) \label{eq:Wconstruction},\\
      B(\theta) &:= \vtf_m(\theta_B),\\
      N(\theta) &:= \vtsu(\theta_N)^\T - \vtsu(\theta_N),\\
      R(\theta) &:= \begin{bmatrix} I_{\dimx} & 0 \end{bmatrix} W(\theta) \begin{bmatrix} I_{\dimx} & 0 \end{bmatrix}^\T,\\
      P(\theta) &:= \begin{bmatrix} I_{\dimx} & 0 \end{bmatrix} W(\theta) \begin{bmatrix} 0 & I_{\dimu} \end{bmatrix}^\T,\\
      S(\theta) &:= \begin{bmatrix} 0 & I_{\dimu} \end{bmatrix} W(\theta) \begin{bmatrix} 0 & I_{\dimu} \end{bmatrix}^\T.
    \end{align}
    \label{eq:PHParamMatrices}
  \end{subequations}
  Then the parametric system
  \begin{align}
    \label{eq:pHParam}
    \pHsys(\theta):
    \begin{cases}
      \!\begin{aligned}
        E(\theta)\dot x(t)= &(J(\theta)-R(\theta))x(t) \\ &\quad+ (B(\theta)-P(\theta)) u(t),
      \end{aligned}\\
      \!\begin{aligned}
        \phantom{E(\theta)}y(t)= &(B(\theta)+P(\theta))^\T x(t)\\ &\quad+ (S(\theta)+N(\theta)) u(t),
      \end{aligned}\\
    \end{cases}
  \end{align}
  satisfies the pH structural constraints. Conversely, for any pH system $\pHsys$ as in Def.~\ref{def:pHDesSys} with $\dimx$ states and $\dimu$ inputs and outputs a vector $\theta \in \R^{n_\theta}$ with $n_\theta = \dimx \left(\frac{3\dimx +1}{2} + 2\dimu\right)+\dimu^2$ can be assigned such that $\pHsys = \pHsys(\theta)$ with $\pHsys(\theta)$ as defined in \eqref{eq:pHParam}.
\end{lem}

\begin{pf}
  For all $v \in \R^{n}$ the terms $\vtu(v)^\T \vtu(v)$ and $\vtsu(v)^\T-\vtsu(v)$ result in positive semi-definite or skew-symmetric matrices, respectively. Therefore, the structural constraints imposed in Def.~\ref{def:pHDesSys} are automatically satisfied when the system matrices are constructed as in \eqref{eq:PHParamMatrices}. This guarantees the pH structure of $\pHsys(\theta)$ for any $\theta \in \R^{n_\theta}$. For the converse statement, a pivoted Cholesky-decomposition of $W$ and $E$, the strictly upper triangular parts of $J$ and $N$, and a vectorization of the $B$ matrix of a system as in Def.~\ref{def:pHDesSys} reveal the appropriate parameter vector.
\end{pf}

\begin{rem}
Note that $\pHsys( \cdot)$ is not an injective map. In fact, for a pH system with dimensions $\dimx=1$ and $\dimu=1$, the two parameter vectors $\theta_1 = \begin{bmatrix}
  1, & 1, & 0, & 1, & 1
\end{bmatrix}^\T
$ and $\theta_2 = \begin{bmatrix}
  -1, & -1, & 0, & 1, & -1
\end{bmatrix}^\T
$ get mapped to the same pH system
\begin{align*}
  \pHsys(\theta_1)=
  \pHsys(\theta_2):
  \begin{cases}
  \dot x(t) = -x(t)+u(t), \\
  y(t) = x(t) + u(t).
  \end{cases}
\end{align*}
\end{rem}

\textcolor{black}{
\begin{rem}
  In Def.~\ref{def:pHDesSys} we only require $E \ge 0$, which allows for singular $E$. In this way, systems with a higher \emph{index} or even singular systems can be expressed with the parametrization in Lemma~\ref{lem:PHParam}. Such systems can cause problems during simulation and control. However, for pH systems, the index is at most two as shown in \cite{Mehl2018}. Furthermore, in \cite{Beattie2016} an index reduction method is introduced, which is again used in \cite{Mehl2018} to remove parts of the system that may impact its stability. Finally, in \cite{Mehl2020} the distance of a given pH system to a potentially \emph{critical} system is studied, such that a test for any harmful properties of identified systems resulting from the singularity of $E$ can easily be implemented. In this way, problems that may occur from requiring only semi-definiteness for the $E$ matrix can be handled effectively.
\end{rem}
}

\subsection{Objective Function and Gradient Computation}

In \cite{SchV20} the goal of obtaining an $\mathcal{H}_\infty$-optimal reduced order model from \emph{exact} FRD leads to an involved bilevel optimization procedure, which focuses the optimization on frequency samples at which the model mismatch is large. On the other hand, for the identification task, a simple least-squares based approach is more suitable, as using the same optimization approach as in \cite{SchV20} would lead to an over-fit to outliers. Therefore, we propose to minimize 
\begin{align}
  \label{eq:objective}
\loss\big(\big(\hat H_i, s_i\big)_{i=1}^{\nSamples}, \tfPHFit( \cdot, \theta)\big) := \sum \limits_{i=1}^{\nSamples} \big\|\hat H_i-\tfPHFit(s_i, \theta)\big\|_2^2,
\end{align}
with respect to the parameter vector $\theta$ in the course of our identification procedure.

To minimize $\loss$, we propose to use a gradient-based optimization algorithm. For a low iteration cost, it is essential to be able to compute a gradient of $\|\hat H_i-\tfPHFit(s_i,\theta)\|_2$ for a sample point $s_i\in \C$ with respect to the parameter vector $\theta$, analytically. This is addressed in the following theorem.

\begin{thm}
  \label{thm:gradients}
  Let $H_0 \in \C^{\dimu \times \dimu}$, $\theta_0 \in \R^{n_\theta}$, and $s_0 \in \C$ be given and assume that $\tfPHFit( s_0,  \cdot): \Omega_{\theta_0}  \rightarrow \C^{\dimu \times \dimu}$, where $\Omega_{\theta_0} \subseteq \R^{n_\theta}$ is a neighborhood of $\theta_0$. Furthermore, assume that the maximal singular value of $H_0-\tfPHFit(s_0, \theta_0)$ is simple and let $\uu, \vv \in \C^\dimu$ be the left and right singular vectors corresponding to the maximum singular value of $H_0-\tfPHFit(s_0, \theta_0)$, respectively. Then $\theta  \mapsto \|H_0-\tfPHFit(s_0, \theta)\|_2$ is differentiable in a neighborhood of $\theta_0$. Moreover, let $\dyn=s_0E(\theta_0)-(J(\theta_0)-R(\theta_0))$ be invertible and define
  \begin{align*}
    \dyninvinp &:= \dyn^{-1}(B(\theta_0)-P(\theta_0))\vv, \\
    \outinvdyn &:= \dyn^{-\H}(B(\theta_0)+P(\theta_0))\uu, \text{ and} \\
    M &:=
    \begin{bmatrix}
      \dyninvinp^{\T} & \vv^{\T}
    \end{bmatrix}^{\T}
    \begin{bmatrix}
      -\outinvdyn^{\H} & \uu^{\H}
    \end{bmatrix}.
  \end{align*}
  Then the gradient $\nabla_\theta \|H_0-\tfPHFit(s_0, \theta_0)\|_2$ is given by $\begin{bmatrix}
    \dE^\T, & \dJ^\T, & \dW^\T, & \dB^\T, & \dN^\T
  \end{bmatrix}^\T
  $, where
  \begin{align*}
    \dE &= -\Real\left(\utv\left(-s_0 \vtu(\theta_E) \left((\dyninvinp \outinvdyn^{\H})^\T +\dyninvinp \outinvdyn^{\H}\right)\right)\right),\\
    \dJ &= -\Real \left( \sutv(-\dyninvinp \outinvdyn^{\H}+(\dyninvinp \outinvdyn^{\H})^\T)\right),\\
    \dW &= -\Real \left( \utv \left( \vtu(\dW) \left(M^\T +M
  \right)\right)\right), \\
          \dB &= -\Real \left( \vectorize(\dyninvinp \uu^{\H} + (\vv \outinvdyn^{\H})^\T) \right),\\
          \dN &= -\Real \left( \sutv(-\vv \uu^{\H} + (\vv \uu^{\H})^\T) \right).
  \end{align*}
\end{thm}

\begin{pf}
  The proofs for $\dE, \dJ, \dB, $ and $\dN$ are analogous to the proof of \cite[Theorem~3.1]{SchV20}. Hence, we only show the result for $\dW$. For that, fix an $i \in \{ \dimx+1, \dots, n_W\}$, where $n_W =\dimx+(\dimx+\dimu)(\dimx+\dimu+1)/2$, and let $e_i$ be the $i$-th standard basis vector in $\R^{n_W}$. Consider the Taylor series expansion of $\varepsilon \mapsto W(\theta_0+\varepsilon e_i)$ at $0$, which is given by
  \begin{align*}
    W(\theta_0+\varepsilon e_i) = W(\theta_0) + \varepsilon \Delta^W_i + \mathcal{O}(\varepsilon^2), 
  \end{align*}
  where $\Delta^W_i := \vtu(\theta_W)^\T \vtu(e_i) + \vtu(e_i)^\T \vtu(\theta_W)$, with $\theta_W$ obtained from $\theta_0$ via the partitioning in Lemma~\ref{lem:PHParam}. Thus, the Taylor series expansion of $\varepsilon \mapsto \tfPHFit(s_0, \theta_0+\varepsilon e_i)$ is given by
  \begin{align*}
    \tfPHFit(s_0, \theta_0+\varepsilon e_i) =& \tfPHFit(s_0, \theta_0) + \\ &\varepsilon
    \begin{bmatrix}
      -\mathcal{C}_0 & I_{\dimu} 
    \end{bmatrix}
    \Delta_i^W
    \begin{bmatrix}
      \mathcal{B}_0^{\T} & I_{\dimu}
    \end{bmatrix}^{\T}
    + \mathcal{O}(\varepsilon^2),
  \end{align*}
  where $\mathcal{B}_0 := \dyn^{-1}\left(B(\theta_0)-P(\theta_0)\right)$ and $\mathcal{C}_0 := (B(\theta_0)+P(\theta_0))^{\T}\dyn^{-1}$. Since the maximum singular value of $H_0 - \tfPHFit(s_0, \theta_0)$ is simple, we have that $\varepsilon \mapsto \| H_0-\tfPHFit(s_0, \theta_0+\varepsilon e_i)\|_2$ is differentiable at zero by \cite{Lan64}. Thus we obtain
  \begin{align*}
    \frac{\dd}{\dd \varepsilon} \| &H_0-\tfPHFit(s_0, \theta_0+\varepsilon e_i\|_2 \Big|_{\varepsilon=0} \\
    &=-\Real\left(\uu^{\H}
    \begin{bmatrix}
      -\mathcal{C}_0 & I_{\dimu}
    \end{bmatrix}
    \Delta_i^W
    \begin{bmatrix}
      \mathcal{B}_0^{\T} & I_{\dimu}
    \end{bmatrix}^{\T}
    \vv
    \right) \\
    &=
    -\Real\left(\trace \left(
    \begin{bmatrix}
      \mathcal{B}_0^{\T} & I_{\dimu}
    \end{bmatrix}^{\T}
    \vv
        \uu^{\H}
    \begin{bmatrix}
      -\mathcal{C}_0 & I_{\dimu}
    \end{bmatrix}
    \Delta_i^W
    \right)
    \right) \\
    &=
    -\Real\Big(\trace \Big(
    \begin{bmatrix}
      \dyninvinp^{\T} & \vv^{\T}
    \end{bmatrix}^{\T}
    \begin{bmatrix}
      -\outinvdyn^{\H} & \uu^{\H}
    \end{bmatrix}
    \Delta_i^W
    \Big)
    \Big) \\
    &=
    -\Real \Big( e_i^\T \utv \Big(\vtu(\dW)(M^{\T} +M) \Big)  \Big).
  \end{align*}
  The last equality is due to \cite[Lemma~3.2]{SchV20}.
\end{pf}

\textcolor{black}{
\begin{rem}
  Note that in Def.~\ref{def:pHDesSys}, the conditions imposed on $J$ and $N$ can be expressed more concisely by introducing a matrix
  \begin{align*}
    \Gamma := \begin{bmatrix}
      J & B \\
      -B^{\T} & N
    \end{bmatrix}
  \end{align*}
  and requiring that $\Gamma^\T = -\Gamma$. This consideration suggests a more concise parametrization of a pH system, in which only $\Gamma$ is parametrized as skew-symmetric matrix and then $J$, $B$, and $N$ are extracted from $\Gamma$ in the same way as $R$, $P$, and $S$ are extracted from $W$ in Lemma~\ref{lem:PHParam}. However, this parametrization is computationally less efficient, since then the fact that there are no constraints on the $B$ matrix, which allows that the gradient can be extracted by a simple vectorization as in Theorem~\ref{thm:gradients}, cannot be not exploited. Preliminary runtime tests indicate that the gradient computation requires around 25\% more computation time when using this alternative parametrization. Therefore, we use the parametrization introduced in Lemma~\ref{lem:PHParam}.
\end{rem}
}

\subsection{Fixed Feedthrough Terms}
\label{sec:fixed_feedthrough}

In our numerical evaluation in Section~\ref{sec:numerical_results}, we observe that despite a good fit in the frequency range covered by the supplied FRD, for large frequencies, the approximation error increases drastically. The cause of this is a bad fit of the feedthrough term. In the method proposed in \cite{Ben2020}, the feedthrough term is assumed to be known and nonzero. Subsequently, the identified model in \cite{Ben2020} is constructed to exactly match that given feedthrough term. However, incorporating this feature into our method is not straightforward. Notice that if we split $\theta_W = \begin{bmatrix}
  \theta_1^\T & \theta_2^\T
\end{bmatrix}^\T
$, where $\theta_1 \in \R^{n_W - m(m-1)/2}$ and $\theta_2 \in \R^{m(m-1)/2}$, then our construction of $W(\theta)$ as in \eqref{eq:Wconstruction} reveals via
\begin{align*}
  \begin{bmatrix}
    W_{11} & W_{12 } \\
    W_{21} & W_{22 } 
  \end{bmatrix}
  &=\vtu\left( \begin{bmatrix}
    \theta_1 \\ \theta_2
\end{bmatrix}\right)^\T
  \vtu\left( \begin{bmatrix}
    \theta_1 \\ \theta_2
\end{bmatrix}\right)\\
  &= 
  \begin{bmatrix}
    \xi_1(\theta_1) & \xi_2(\theta_1) \\
    0 & \xi_3(\theta_2) \\
  \end{bmatrix}^\T
  \begin{bmatrix}
    \xi_1(\theta_1) & \xi_2(\theta_1) \\
    0 & \xi_3(\theta_2) \\
  \end{bmatrix},
\end{align*}
where $\xi_1$, $\xi_2$, and $\xi_3$ arrange $\theta_1$ and $\theta_2$ into the Cholesky factors of $W(\theta)$,
that $S(\theta)$ is dependent on both $\theta_1$ and $\theta_2$. Therefore, to fix $S$, a constraint on $\theta_1$ and $\theta_2$ must be imposed in every optimization step. This can be circumvented by an alternative definition of $W(\theta)$ via $\widetilde W(\theta) =\vtu(\theta_W)\vtu(\theta_W)^\T$, since in this way $\widetilde S(\theta)= \begin{bmatrix} 0 & I_{\dimu} \end{bmatrix} \widetilde W(\theta) \begin{bmatrix} 0 & I_{\dimu} \end{bmatrix}^\T$ only depends on $\theta_2$, which can then simply be fixed to the desired value.

However, another subtlety that must be handled when simply fixing the symmetric part of the feedthrough term using $\widetilde W(\theta)$ is the implicit constraint that is in turn imposed on the $P$ and $R$ matrices of the pH system. Via the Schur complement the expression $W\ge 0$ in Def.~\ref{def:pHDesSys} can be reformulated (see \cite{Horn2005}) as
\begin{align}
  R \ge 0,\; S-P^\T R^+ P \ge 0,\; (I-R R^+)P = 0,
\end{align}
where $R^+$ denotes the Moore-Penrose pseudoinverse of $R$, such that by fixing $S$, we obtain an implicit constraint involving the pseudoinverse of $R$. In Section~\ref{sec:numerical_results}, we can observe that this obstructs the parameter optimization.

For this reason, we propose to implement the symmetric part of a fixed feedthrough term as a soft constraint, i.e.\ add a penalty term $\lambda {\left\|S(\theta)-S_{\text{given}}\right\|}_2^2$ to the objective function \eqref{eq:objective} to obtain
\begin{align}
  \label{eq:losslambda}
  \begin{split}
  \loss_\lambda\big(\big(\hat H_i, s_i\big)_{i=1}^{\nSamples}, \tfPHFit( \cdot, \theta), S_{\text{given}}\big) := \phantom{tttttesttestte} \\
  \loss\big(\big(\hat H_i, s_i\big)_{i=1}^{\nSamples}, \tfPHFit( \cdot, \theta)\big) + \lambda \big\| S(\theta) - S_{\text{given}}\big\|_2^2.
  \end{split}
\end{align}
The parameter $\lambda$ can either be chosen beforehand or be dynamically adjusted during the optimization. Note that, on the other hand, the skew-symmetric part of the feedthrough term can simply be fixed to some skew-symmetric matrix $N_{\text{given}}$, since this does not lead to any implicit constraints on other parts of the system.

\section{Numerical Experiments}
\label{sec:numerical_results}

We assess the accuracy of the identified models resulting from our method in a numerical comparison with pH-Loewner\footnote{sourcecode available at \url{github.com/goyalpike/Identify\_PortHamiltonian\_Realization}} and the vector-fitting framework\footnote{sourcecode available at \url{github.com/pedrohnv/vectfit\_julia/}} on a 200 dimensional electrical circuit benchmark system presented in \cite{Gug03}. This is also used in \cite{Ben2020} to highlight the effectiveness of the pH-Loewner approach.

\subsection{Experiment Setup}

\begin{figure}
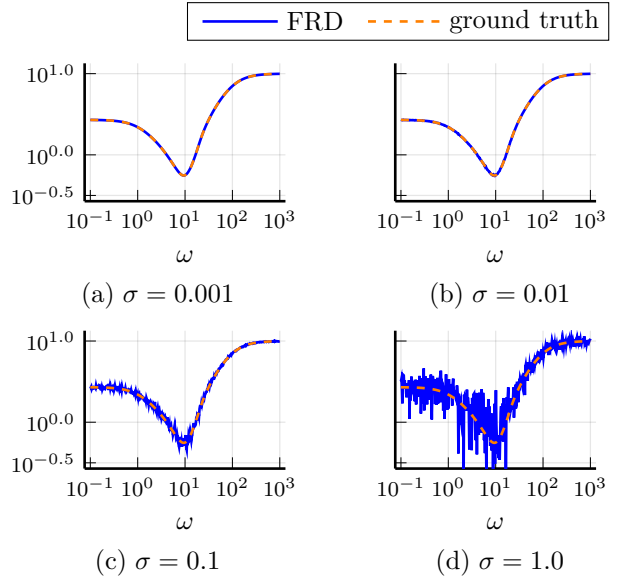

  \hfill
    \begin{tikzpicture}
      \draw (-0.2, -0.25) rectangle (5.4,0.25);
      \draw [-, blue, line width=1.0pt] (0,0) -- (1,0);
      \node at (2,0)  [anchor=east] {FRD};
      \draw [-, orange, dashed, line width=1.0pt] (2.2,0) -- (3.2,0);
      \node at (5.4,0) [anchor=east] {ground truth};
    \end{tikzpicture} \\
    \begin{tabular}{cc}
      \input{"PlotSources/sigma1em3.tex"} &
      \input{"PlotSources/sigma1em2.tex"} \vspace{-0.2cm}\\
      (a) $\sigma=0.001$ & (b) $\sigma=0.01$\\
      \input{"PlotSources/sigma1em1.tex"} &
  \input{"PlotSources/sigma1em0.tex"} \vspace{-0.2cm}\\
      (c) $\sigma=0.1$ & (d) $\sigma=1.0$
    \end{tabular}
  \caption{Transfer function of ground truth model and samples FRD used for identification at different noise levels $\sigma$. 
  }
  \label{fig:groundtruth}
\end{figure}

Our test setup is as follows: As in \cite{Ben2020} we collect transfer function evaluations of the benchmark system at 400 logarithmically spaced points on the imaginary axis between $10^{-2}$ and $10^{1}$. Then we perturb the evaluations with zero mean Gaussian noise with standard deviations of $10^{-3}, 10^{-2}, 10^{-1}$, and $10^0$. For each standard deviation, we compute 20 sets of perturbed frequency samples. In Figure~\ref{fig:groundtruth}, the transfer function of the benchmark is depicted together with four samples of FRD (one for each considered standard deviation) that are later used for identification. Since in vector-fitting and in our method, the model order of the identified model can be chosen beforehand, we identify models with orders 3, 5, 7, and 9 on all 80 datasets. The model order of the resulting pH-Loewner models cannot be set beforehand and is around 400, since the reduction of the Loewner pencils fails in presence of noisy data.

We evaluate three different variants of our proposed method which handle the feedthrough term differently. First, we test our identification method without any attention to a potentially known feedthrough term, such that we just minimize the objective function given in \eqref{eq:objective} to obtain a parametrization. The second variant considers a known feedthrough term and fixes the feedthrough of the identified model using the alternative definition of $W(\theta)$ as explained in Section~\ref{sec:fixed_feedthrough}. The last variant uses the extra feedthrough penalty term as in the extended objective function in \eqref{eq:losslambda} to allow for an adjustment of the feedthrough of the identified model. The tuning parameter $\lambda$ in the penalty term is chosen equal to the respective noise level $\sigma$. We denote these three approaches by pH-flex, pH-fixed, and pH-reg, respectively. In all three variants, we use the Broyden-Fletcher-Goldfarb-Shanno (BFGS) quasi-Newton optimization method with linesearch based on the Strong-Wolfe conditions. We use the implementation described in \cite{mogensen2018optim}.

The initialization options chosen for the different identification methods are as follows. In vector-fitting, we set the initial poles of the to-be-fitted transfer function logarithmically spaced on the negative real axis as is common practice in vector-fitting applications. Vector-fitting allows to weight different frequencies to achieve a higher accuracy at desired samples. In our experiments we set all initial weights to 1. We enforce the stability of the system in every iteration step and constrain $\vfE$ to zero to obtain a proper transfer function. Furthermore, we employ the \emph{relaxed nontriviality constraint} as proposed in \cite{Gus06}. We initialize pH-flex and pH-reg with a pseudo-random parameter vector; pH-fixed is initialized in the same way, except for the elements of the parameter vector that impact the feedthrough-term, which are set to match the given feedthrough and are fixed during optimization. The pH-Loewner method does not require any initialization options.

\subsection{Experimental Results}

We present our experimental results in the following order. We start with an accuracy comparison between pH-Loewner, vector-fitting, and our method. After that, we study the effects of using the different variants of our method more carefully. Finally, we compare the influence of the different predefined model orders on the accuracy of the identified models for different noise levels. 

\begin{figure}
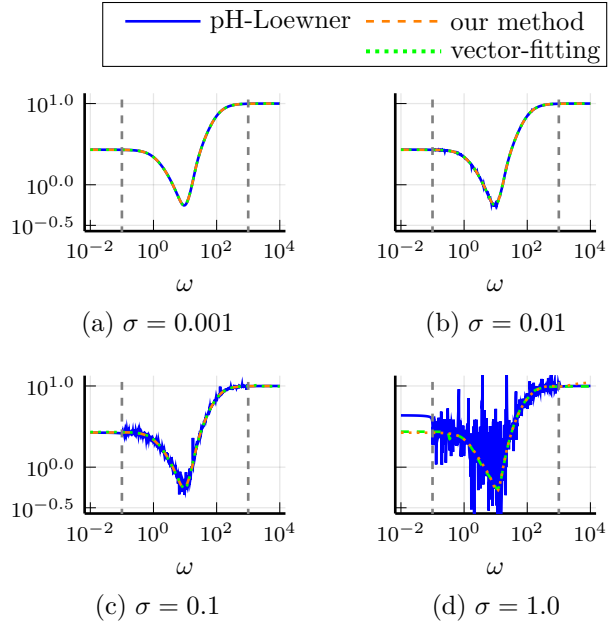

  \hfill
    \begin{tikzpicture}
      \draw (-0.3, -0.6) rectangle (6.4, 0.25);
      \draw [-, blue, line width=1.0pt] (0,0) -- (1,0);
      \node at (1,0) [anchor=west] {pH-Loewner};
      \draw [-, orange, dashed, line width=1.0pt] (3.2,0) -- (4.2,0);
      \node at (4.2,0) [anchor=west] {our method};
      \draw [-, green, dotted, line width=1.5pt] (3.2,-0.4) -- (4.2,-0.4);
      \node at (4.2,-0.4) [anchor=west] {vector-fitting};
    \end{tikzpicture} \\
  \begin{tabular}{cc}
    \input{"PlotSources/samples-0.001.tex"} & \input{"PlotSources/samples-0.01.tex"} \vspace{-0.2cm} \\
    (a) $\sigma=0.001$ & (b) $\sigma=0.01$ \vspace{0.2cm} \\
    \input{"PlotSources/samples-0.1.tex"} & \input{"PlotSources/samples-1.0.tex"} \vspace{-0.2cm} \\
    (c) $\sigma=0.1$ & (d) $\sigma=1.0$ \vspace{0.2cm} \\
  \end{tabular}
  \caption{Identified transfer functions for different noise levels $\sigma$. 
  The frequency range in which the provided FRD is located is bounded by gray dashed lines.}
  \label{fig:sample_identification}
\end{figure}

In Figure~\ref{fig:sample_identification}, samples of identified transfer functions are shown for different noise levels $\sigma$. It can be observed that for small $\sigma$ all three methods approximate the main features of the given transfer function. However, for larger noise levels, it becomes obvious that pH-Loewner rather augments the error at certain frequencies, which can be seen by comparing the amplitude of the peaks of the pH-Loewner transfer function in Figure~\ref{fig:sample_identification} (c-d), with the peaks of the corrupted sample data in Figure~\ref{fig:groundtruth} (c-d). On the other hand, the compensatory properties of vector-fitting and our method lead to good approximations even for larger $\sigma$.

\begin{figure}
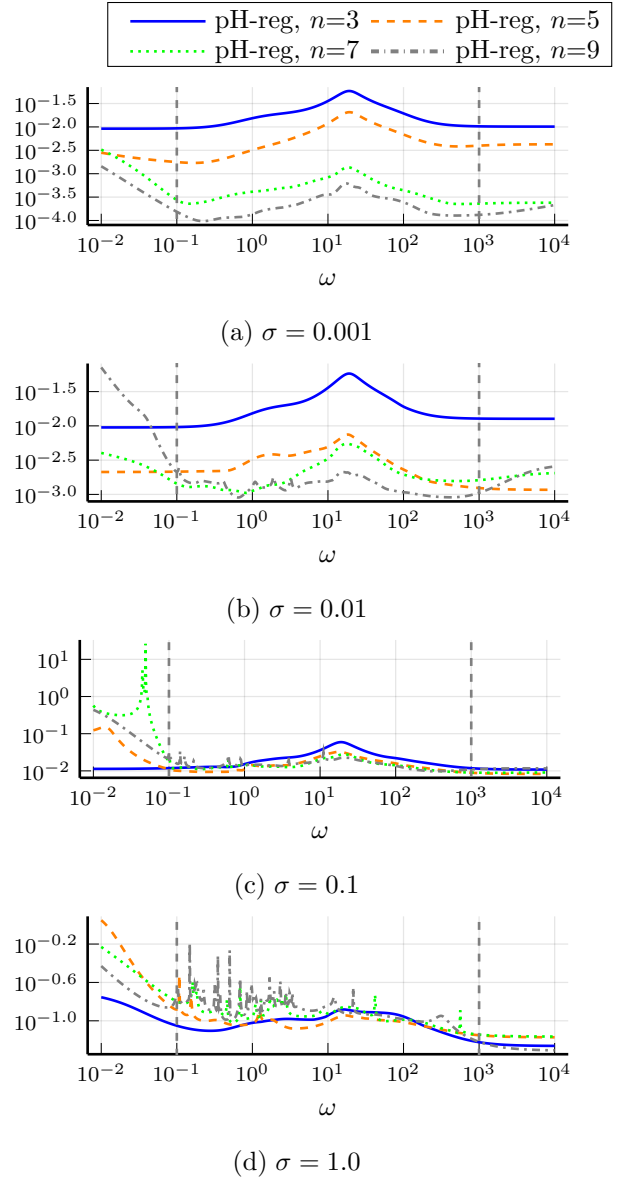

  \hfill
    \begin{tikzpicture}
      \draw (-0.3, -0.6) rectangle (6.4, 0.25);
      \draw [-, blue, line width=1.0pt] (0,0) -- (1,0);
      \node at (1,0) [anchor=west] {pH-reg, $n$=3};
      \draw [-, green, dotted, line width=1.0pt] (0,-0.4) -- (1,-0.4);
      \node at (1,-0.4) [anchor=west] {pH-reg, $n$=7};
      \draw [-, orange, dashed, line width=1.0pt] (3.2,0) -- (4.2,0);
      \node at (4.2,0) [anchor=west] {pH-reg, $n$=5};
      \draw [-, gray, dashdotted, line width=1.5pt] (3.2,-0.4) -- (4.2,-0.4);
      \node at (4.2,-0.4) [anchor=west] {pH-reg, $n$=9};
    \end{tikzpicture} \\
  \begin{tabular}{c}
    \input{"PlotSources/differentOrdersErrors/different-orders-0.001.tex"} \\
    (a) $\sigma=0.001$\\
    \input{"PlotSources/differentOrdersErrors/different-orders-0.01.tex"} \\
    (b) $\sigma=0.01$ \\
    \input{"PlotSources/differentOrdersErrors/different-orders-0.1.tex"} \\
    (c) $\sigma=0.1$ \\
    \input{"PlotSources/differentOrdersErrors/different-orders-1.0.tex"} \\
    (d) $\sigma=1.0$ \\
  \end{tabular}
  \caption{Comparison of mean errors of pH-reg for varying model orders at different noise levels. 
  The frequency range in which the provided FRD is located is bounded by gray dashed vertical lines.}
  \label{fig:differentorders}
\end{figure}

We study the approximation error in more detail and measure it using the error function
\begin{align}
  \label{eq:err_fun}
  \errmeasure \left(H_{\mathsf{true}}, H_{\mathsf{id}}, S_v\right) := \frac{1}{n_{S_v}}\sum\limits_{s_i \in S_v} {\left\|H_{\mathsf{true}}(s_i)-H_{\mathsf{id}}(s_i)\right\|}_2,
\end{align}
where $S_{v}$ is a set of $n_{S_v}=900$ logarithmically spaced validation sample points between $10^{-2}$ and $10^{1}$, which has an empty intersection with the set of (training) sample points used during identification but lies in the same frequency range. In Tab.~\ref{tab:acc_comp} the mean of $\errmeasure$ over the 20 identified models per identification algorithm is reported for varying noise levels. For brevity, we only report the errors of vector-fitting and the variants of our method for predefined model orders of 9 and defer reviewing the effect of different model orders to Figure~\ref{fig:differentorders}. Note that for each noise level our method leads to the lowest mean error. We conduct a statistical analysis and a paired sample t-test reveals that for each noise level our method (variants pH-flex and pH-reg) leads to models with a significantly lower error than the models obtained with pH-Loewner with p-values of less than $0.02$. In comparison to vector-fitting, pH-flex and pH-reg lead to significantly more accurate models for noise levels $\sigma \le 0.1$ (with p-values less than $0.02$). The significance level rises as $\sigma$ is reduced.

\begin{rem}
  The large mean error of vector-fitting at the noise level $\sigma=1$ is caused by three outliers, with errors of $1.56$, $2.85$, and $23.58$, respectively. The mean accuracy in terms of $\errmeasure$ of the remaining models for samples at $\sigma=1$ is comparable to our method. This is why the t-test does not reveal a significant increase in accuracy at that noise level despite the large difference of the mean values.
\end{rem}

\begin{table}
  \caption{Accuracy comparison for different identification methods and noise levels. The lowest mean error is highlighted in gray. Vector-fitting is denoted by vectfit.}
  \label{tab:acc_comp}
  \centering
  \begin{tabular}{c|cccccc}
    Method    & $\dimx$    & $\sigma$& mean error  & std.\ deviation \\ \hline 
    Loewner   & 400  & 0.001  & 1.03e$-$3  & 1.48e$-$4 \\ 
    vectfit   & 9    & 0.001  & 4.16e$-$4  & 1.02e$-$5 \\ 
    pH-flex   & 9    & 0.001  & 2.38e$-$4  & 1.11e$-$4 \\ 
    pH-fixed  & 9    & 0.001  & 1.42e$-$2  & 2.26e$-$2 \\ 
    pH-reg    & 9    & 0.001  & \cellcolor{black!25} 2.23e$-$4  & 1.12e$-$4 \\ 
    \hline
    Loewner   & 400  & 0.01  & 1.00e$-$2  & 6.15e$-$4 \\ 
    vectfit   & 9    & 0.01  & 3.36e$-$3  & 3.29e$-$4 \\
    pH-flex   & 9    & 0.01  & \cellcolor{black!25} 1.32e$-$3  & 2.45e$-$4 \\ 
    pH-fixed  & 9    & 0.01  & 1.39e$-$2  & 2.27e$-$2 \\ 
    pH-reg    & 9    & 0.01  & 1.34e$-$3  & 2.59e$-$4 \\ 
    \hline
    Loewner   & 400  & 0.1  & 1.06e$-$1  & 1.16e$-$2 \\ 
    vectfit   & 9    & 0.1  & 1.74e$-$2  & 1.70e$-$3 \\ 
    pH-flex   & 9    & 0.1  & \cellcolor{black!25} 1.46e$-$2  & 3.69e$-$3 \\ 
    pH-fixed  & 9    & 0.1  & 2.22e$-$1  & 8.25e$-$1 \\ 
    pH-reg    & 9    & 0.1  & 1.50e$-$2  & 3.86e$-$3 \\ 
    \hline
    Loewner   & 400  & 1.0  & 3.10e$+$0  & 4.60e$+$0 \\ 
    vectfit   & 9    & 1.0  & 1.49e$+$0  & 5.25e$+$0 \\
    pH-flex   & 9    & 1.0  & 1.20e$-$1  & 2.39e$-$2 \\ 
    pH-fixed  & 9    & 1.0  & \cellcolor{black!25} 1.10e$-$1  & 2.29e$-$2 \\ 
    pH-reg    & 9    & 1.0  & 1.30e$-$1  & 1.93e$-$2 \\ 
  \end{tabular}
\end{table}

\begin{figure}
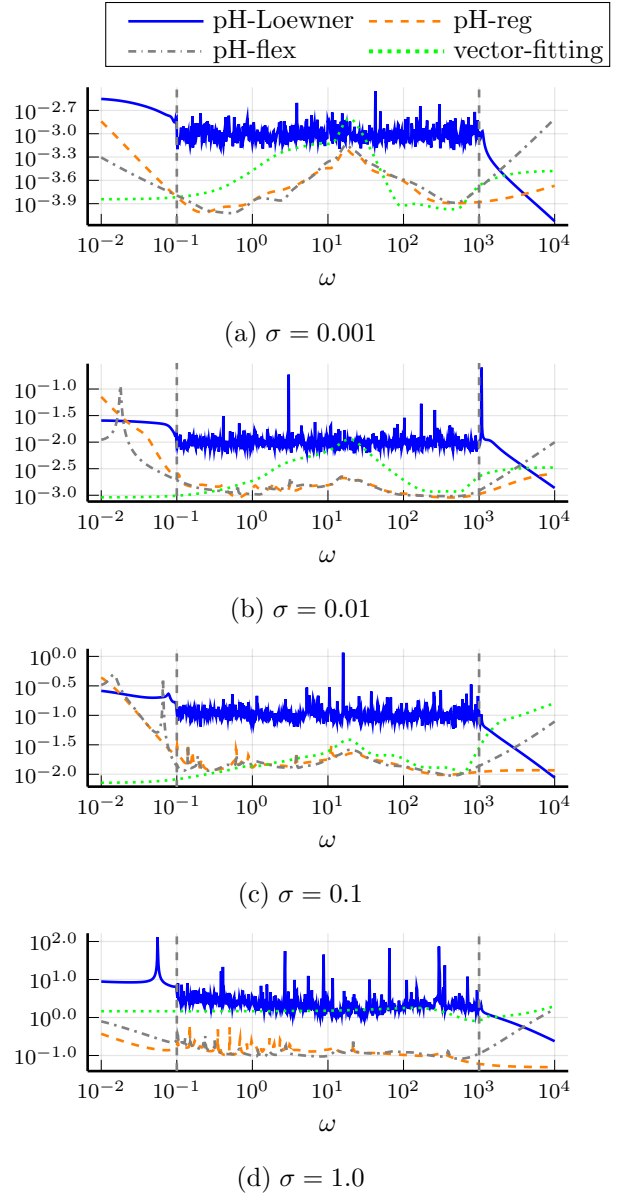

  \hfill
    \begin{tikzpicture}
      \draw (-0.3, -0.6) rectangle (6.4, 0.25);
      \draw [-, blue, line width=1.0pt] (0,0) -- (1,0);
      \node at (1,0) [anchor=west] {pH-Loewner};
      \draw [-, gray, dashdotted, line width=1.0pt] (0,-0.4) -- (1,-0.4);
      \node at (1,-0.4) [anchor=west] {pH-flex};
      \draw [-, orange, dashed, line width=1.0pt] (3.2,0) -- (4.2,0);
      \node at (4.2,0) [anchor=west] {pH-reg};
      \draw [-, green, dotted, line width=1.5pt] (3.2,-0.4) -- (4.2,-0.4);
      \node at (4.2,-0.4) [anchor=west] {vector-fitting};
    \end{tikzpicture} \
  \begin{tabular}{c}
    \input{"PlotSources/meanErrors/meanError0.001.tex"} \vspace{-0.0cm}\\
    (a) $\sigma=0.001$ \\
    \input{"PlotSources/meanErrors/meanError0.01.tex"}  \vspace{-0.0cm} \\
    (b) $\sigma=0.01$ \\
    \input{"PlotSources/meanErrors/meanError0.1.tex"}   \vspace{-0.0cm} \\ 
    (c) $\sigma=0.1$ \\
    \input{"PlotSources/meanErrors/meanError1.0.tex"}   \vspace{-0.0cm} \\
    (d) $\sigma=1.0$
  \end{tabular}
  \caption{Mean errors of identified transfer functions for varying noise levels and different identification methods. 
  The frequency range in which the provided FRD is located is bounded by gray dashed vertical lines.}
  \label{fig:mean_errors}
\end{figure}

A frequency-wise comparison of the mean errors (across the 20 frequency sample sets in each noise level) is shown in Figure~\ref{fig:mean_errors}. The mean errors of the pH-Loewner models are approximately the same as the given noise level, while vector-fitting and our method have mean errors almost an order of magnitude below the noise level over a wide frequency range, since these two methods compensate noise. Looking at the mean error at frequencies outside the frequency range in which FRD is available, we note that for low frequencies, our method leads to larger errors, while vector-fitting maintains a low error also at frequencies well below the given FRD range. However, for high frequencies we can observe that the regularization used in pH-reg leads to a better fit even beyond the given FRD range.

\begin{figure}
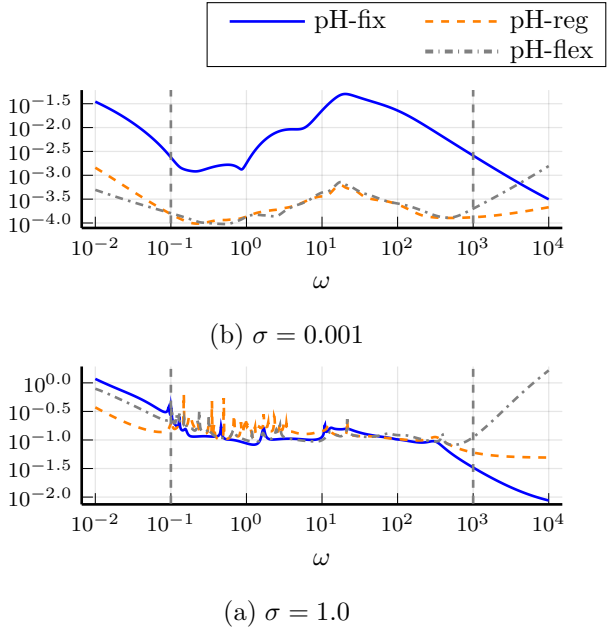

  \hfill
    \begin{tikzpicture}
      \draw (-0.3, -0.6) rectangle (5.0, 0.25);
      \draw [-, blue, line width=1.0pt] (0,0) -- (1,0);
      \node at (1,0) [anchor=west] {pH-fix};
      \draw [-, orange, dashed, line width=1.0pt] (2.6,0) -- (3.6,0);
      \node at (3.6,0) [anchor=west] {pH-reg};
      \draw [-, gray, dashdotted, line width=1.5pt] (2.6,-0.4) -- (3.6,-0.4);
      \node at (3.6,-0.4) [anchor=west] {pH-flex};
    \end{tikzpicture} \\
  \begin{tabular}{c}
    \input{"PlotSources/variantErrors/variants0.001.tex"}\\ 
    (b) $\sigma=0.001$ \\
    \input{"PlotSources/variantErrors/variants1.0.tex"}\\
    (a) $\sigma=1.0$ \\
  \end{tabular}
  \caption{Comparison of method variants for different noise levels.
  The frequency range in which the provided FRD is located is bounded by gray dashed vertical lines.}
  \label{fig:variants}
\end{figure}

In Figure~\ref{fig:variants}, the different variants of our methods for incorporating the feedthrough term are compared following up on our discussion in Section~\ref{sec:fixed_feedthrough}. For pH-flex the given term is not considered during the optimization which results in an increased error for large frequencies. On the other hand, for pH-fix the parametrization is chosen such that the feedthrough of the identified model is fixed to the given feedthrough term using the alternative parametrization via $\widetilde W(\theta)$ and $\widetilde S(\theta)$ as in Section~\ref{sec:fixed_feedthrough}. This leads to a fast decay of the error for large frequencies. However, especially for small noise levels, the error at all other frequencies is several orders of magnitudes higher. This is due to the implicit constraint imposed on the other parts of the parameter vector as discussed in Section~\ref{sec:fixed_feedthrough}, which has a negative impact on the optimization.
This implicit constraint is circumvented in the pH-reg models. Figure~\ref{fig:variants} reveals that this approach leads to a better fit for high frequencies without reducing the accuracy in the range of the given FRD.

The accuracy for different predefined model orders is compared in Figure~\ref{fig:differentorders}. As expected, for the smaller noise levels, the fit is worse for the lower order models, since the FRD cannot be captured accurately with the few parameters that the low order models possess. On the other hand, for the largest noise level the lowest model order leads to the best fit, since it is less prone to over-fitting the corrupted data due to the fewer parameters. Furthermore, for the lower order models, the error does not grow as rapidly in the unknown frequency range, while the larger order models exhibit typical signs of over-fitting in this case such as a less smooth error profile and, in particular, the worse fit in the frequency bands lower and larger than the given FRD.

\textcolor{black}{
\begin{rem}
  We also briefly report the resuts of additional experiments, in which $E$ is fixed to the identity matrix. In this way, a pH system consisting of an ordinary differential equation is identified. These experiments indicate that such a parametrization leads to models with a similar accuracy as models with a flexible (and optimized) $E$ matrix. The similar accuracy suggests that our method can also be applied when the resulting identified models are required to be available in standard state-space form.
\end{rem}
}

\section{Conclusion}

We have presented a new method for the identification of passive models in pH format from noisy FRD. The method uses direct parameter optimization to minimize the least-squares error between the given FRD and the transfer function evaluations of the fitted model. The parametrization is chosen such that the resulting identified system is passive for all parameter vectors. We have evaluated the accuracy of the identified models in comparison to pH-Loewner and vector-fitting. The comparison has shown that significantly more accurate models can be obtained when using our new method. Furthermore, we have conducted experiments with different model orders, which reveal the necessity for choosing an appropriate model order depending on the noise level, and have addressed a subtlety that arises when fixing the feedthrough term of the identified model.

While the improved accuracy compared to pH-Loewner was expected, we have noticed the smaller yet still statistically significant increase in accuracy when using our new method in comparison to vector-fitting with great interest. Future research studies whether this is simply due to the more flexible parametrization or if the passive prior of our model causes the increased accuracy, since the underlying model from which the FRD is obtained is also passive.

\section*{Acknowledgment}
I thank Volker Mehrmann for many fruitful discussions on pH systems and his valuable remarks concerning the presented work. Furthermore, I gratefully acknowledge Benjamin Unger and Matthias Voigt for reviewing early versions of this manuscript.

\bibliographystyle{IFACh}
\bibliography{references}
 	





\end{document}